\documentclass[11pt]{article}
\usepackage{graphicx}%
\usepackage{multirow}%
\usepackage{amsmath,amssymb,amsfonts}%
\usepackage{amsthm}%
\usepackage{mathrsfs}%
\usepackage[title]{appendix}%
\usepackage{xcolor}%
\usepackage{textcomp}%
\usepackage{manyfoot}%
\usepackage{booktabs}%
\usepackage{algorithm}%
\usepackage{algorithmicx}%
\usepackage{algpseudocode}%
\usepackage{listings}%

\title{Superposition Principle in Relativistic Gravity}

\author{ Y. Friedman  \\
Extended Relativity Research Center\\
Jerusalem College of Technology, Israel\\
P.O.B. 16031, Jerusalem 91160, Israel\\
e-mail: friedman@g.jct.ac.il}
\begin{document}
\date{}

\maketitle
\bigskip
\noindent

{\small{\bf ABSTRACT.} 
We present a simple model for relativistic gravity. The model represents a gravitational field  by a metric on a Minkowski space background. We introduce a new Lorentz-covariant metric for a single-source gravitation field, which is determined by the retarded position and the velocity of the source. This metric passes the $GR$ tests and properly describes the strong gravitational field. Using this metric, we define the metric of a field generated by several moving sources. The explicit true acceleration of an object moving in this field is derived from the ``first-order" (in the masses of the sources) acceleration, which is linear in the sources. The field splits into near and far fields. We compare the far gravitational field of a binary star with electromagnetic quadruple radiation and present an algorithm for computing accelerations in such a field. These results may lead to a new understanding of gravitational waves and the structure and dynamics of galaxies.}

 
\maketitle
\section{Introduction}
$\;\;\;\;\;$
The most common source of gravitational fields is a collection of moving, spherically symmetric bodies. The superposition principle in Newtonian gravity states that the acceleration of an object in such a field is the sum of the accelerations due to each source. However, there is no known analog of the superposition principle in relativistic gravity -- until now. Here, we present a superposition principle for relativistic gravity which extends the Newtonian one.

A. Einstein's General Relativity ($GR$), introduced in 1915, is the accepted theory of relativistic gravitation.  In $GR$, the gravitational force curves spacetime, and this curving is expressed by a metric \cite{MTW}-\cite{Kop}. The theory is generally covariant. Einstein's non-linear field equations are used to determine the metric from the sources of the field. However, these are complicated equations and can be solved explicitly only for simple cases. For combined fields, the post-Newtonian approximation is used.

In \cite{Bel}, the authors obtain a description of the gravitational field and the equation of motion for two gravitationally interacting bodies in the post-linear approximation of $GR$. In their treatment, they use the Minkowskian parametrization of the worldline, a flat retarded propagator and Poincar\'{e} invariance of the gravitational field. All of these ideas are also essential in the approach presented here. The authors of \cite{Bel} use an approximation scheme, introducing the deviation $h$
of the gravitational field metric  from the Minkowski metric. Next, in order to combine the fields of the two bodies, they decompose $h$ into a power series with respect to mass of the source $M$ (or, in their notation, with respect to the gravitational constant $G$). For example, if one uses the Schwarzschild metric for the single-source gravitational field, the above power series expansion of $h$ is infinite.

However, if one uses a metric for which $h$ is proportional to $M$ for the single-source gravitational field, the superposition becomes much simpler. A metric of this type on Minkowski spacetime, defined by the retarded position of the source, was introduced by Whitehead \cite{White}. It was shown by Eddington \cite{Edding} that for the static case, this metric satisfies Einstein's field equations. The metric was extended by Synge \cite{Synge} to multiple-source gravitational fields. However, a single-source metric of this type does not predicts black holes, has non-natural predictions for the speed of light and has predictions contradicting observations. The predictions of this metric are criticized in \cite{Will08}. Similar metrics were used by Finkelstein \cite{Fink} to show that the Schwarzschild radius is only a coordinate singularity and not a true physical singularity. This metric is based on what are now called \cite{HEL} advanced Eddington-Finketstein coordinates and predicts the correct behavior near black holes. However, the metric is derived using the advanced position of the source, thereby contradicting causality.  

Looking for all possible metrics on a Minkowski space background for a static, spherically symmetric gravitational field, and requiring that the metrics pass the classical tests of $GR$, we found \cite{FStav} a metric for which the deviation $h$ is proportional to $M$. This metric is similar Whitehead's metric, but we added a space reflection. The resulting metric is of the Kerr-Schild type \cite{KerrSchild} and satisfies Einstein's field equations. Moreover, this metric has no singularity at the Schwarzschild radius. In \cite{Adler}, such a metric is called a {\em degenerate} metric. Kerr \cite{Kerr} also used this type metric to obtain an exact solution of Einstein's field equations for a rotating, axially-symmetric black hole, see also \cite{Adler}.

The metrics from \cite{FStav}  appear again in the recent article \cite{FUnify} and the book \cite{ana}, which introduce \emph{Extended Relativity} ($ER$), a new model whose main goal is to incorporate both electromagnetism and gravitation in a unified relativistic theory. In $ER$, the gravitational field of a single, spherically symmetric source determines a retarded four-potential, and the metric defined by the field is degenerate.

In this paper, we introduce a new Lorentz-covariant metric for a gravitational field of a single, spherically symmetric source for which $h$ is proportional to the mass of the source. For a static source, this metric coincides with the metric from \cite{FStav} and \cite{ana}. Since the classical $GR$ tests involve  static sources, this new metric passes these tests. Based on this metric, we define the metric of a gravitational field of a collection of spherically symmetric, moving sources by defining the deviation $h$ of the combined field as a linear combination of the single-source deviations. We also find the inverse of this metric. To derive an explicit formula for the acceleration of objects moving in the field, we use the source-independent Minkowskian parameter $\tau$ (the proper time of the moving object) to parameterize the object's worldline. We observe first that the dynamics equation for motion with respect to $\tau$ in an arbitrary field is determined by the Christoffel symbols of the field. This symbol can be expanded in a power series in the masses of the sources. The first-order term is defined by the first-order tensor of the field and defines the first-order acceleration caused by the field. The first-order acceleration of the combined field is the sum of the first-order accelerations of each source. This is the extension of the Newtonian superposition principle to relativistic gravity. The first-order acceleration explicitly defines the full acceleration caused by the field. By use of the formulas for the derivatives of the retarded position and the retarded velocity, we obtain an explicit form of the first-order tensor. We discover that, like in electromagnetism, the gravitational field splits into a near field and a far field. We compare the gravitational and electromagnetic fields and present an algorithm for computing accelerations in a gravitational field generated by several moving sources.  

\section{The metric of the gravitational field of a spherically symmetric single-source }
$\;\;\;\;\;$
We represent a gravitational field by a metric on the Minkowski spacetime of a very distant observer. The worldline of a test object in this field is a geodesic with respect to the metric. We denote the Minkowski metric tensor by $\eta_{\mu\nu}=diag(1,-1,-1,-1)$.  Raising and lowering of indices in this paper is done via $\eta$. We require only Lorentz covariance, meaning that tensors need obey the tensor transformation rule only with respect to the Poincar\'{e} transformations on Minkowski space. The Minkowskian, Lorentz-invariant parameter $\tau$ is defined by
\[d\tau ^2=\eta_{\mu\nu}dx^\mu dx^\nu .\]
Denote the four-velocity $dx/d\tau$ of the object by $\dot{x}$. The four-velocity is of norm one, that is, $\dot{x}\cdot \dot{x}=1$, where $\cdot$ denotes the Minkowski inner product. This implies that the four-acceleration is orthogonal to the four-velocity.  

Consider now a gravitational field generated by a  spherically symmetric mass $m$. Denote by  
\begin{equation}\label{mass}
 M=Gm/c^2,   
\end{equation} 
where $G$ is the gravitational constant. Thus, $M$ represents the mass of the source in dimensions of length. The worldline that the source is moving along is denoted by $\psi(\tau)$. Given an object at the spacetime point $x$, define the {\em retarded time} $\tau(x)$  as the unique solution of $(x-\psi (\tau))^2=0$ and $x^0-\psi^0 (\tau(x))>0$. The point with coordinates $\psi (\tau(x))$, called the {\em retarded position}, is the unique intersection of the backward light cone with vertex $x$ and the worldline $\psi(\tau)$ of the source. Since the field propagates with the speed of light, the source at the retarded time is the only influence on the object at $x$. The four-velocity and the four-acceleration of the  source at the retarded time $\tau(x)$ will be denoted by $w(x)$ and $a(x)$, respectively.

The {\em relative position} of $x$ with respect to the retarded position of the  source is the null four-vector $r(x)=x-\psi (\tau(x))$. To define the new metric, we have to introduce a conjugation $^{*u}$  of  four-vectors with respect to a norm-one vector $u$, defined by \begin{equation}\label{defConj}
( x^{*u})=2(x\cdot u)u-x
\end{equation}
for any spacetime position $x\in M$. It can be verified directly that $(^{*u})^2=id$ and that this conjugation reverses the spatial direction of the four-covector $r(x)$ in the frame moving with four-velocity $u$, while keeping the time component unchanged. 

With this notation, we introduce a  {\em four-potential of a single source field} 
\begin{equation}\label{lWhit0}
l(x)=\frac{r^{*w}(x)}{(r(x)\cdot w(x))^{3/2}}=\frac{2(r(x)\cdot w(x))w(x)-r(x)}{(r(x)\cdot w(x))^{3/2}}\,.
\end{equation}
Since $r(x)\cdot w(x)$ is Lorentz invariant, any power of it is also Lorentz invariant. In addition, $w(x)$ and $r(x)$ are four-vectors and thus Lorentz covariant. Thus, the covector $l(x)$, defined by \eqref{lWhit0}, is Lorentz covariant. This covector is null. It is the gravitational analog of the Li\'{e}nard-Wiechert four-potential of single source electromagnetic field.  We define the {\em gravitational potential tensor} by
\begin{equation}\label{poteGrav}
    h_{\alpha\beta}(x)=2M l_{\alpha} (x)l_{\beta} (x),
\end{equation}
where the notation $l_{\alpha}$ means the $\alpha$ component of the four-covector $l$. The gravitational potential tensor represents the deviation of the gravitational field metric from the flat space metric and expresses the effect of the gravitational field on flat spacetime. This potential is a Lorentz-covariant tensor of rank $(0,2).$

The new metric of the gravitational field of a single, spherically symmetric source is
\begin{equation}\label{metricNew}
    g_{\alpha\beta}(x)= \eta_{\alpha\beta}- h_{\alpha\beta}(x),
\end{equation}
for $h_{\alpha\beta}(x)$, defined by \eqref{poteGrav}.

In \cite{FStav} and \cite{ana}, we we used this metric for a spherically symmetric source at rest to show that it passes all classical tests of $GR$ and satisfies Einstein's field equation. Note that the deviation of this metric from the Minkowski metric is proportional to the mass $M$ of the source. 

\section{The metric of a multiple-source gravitational field}
$\;\;\;\;\;$
Consider now a gravitational field generated by a collection of spherically symmetric masses $M_j$ moving along wordlines $\psi_j(\tau)$,  for a set of indices $\{j\}$. Given an object at the spacetime point $x$ in this field, define, as above, the  retarded time $\tau_j(x)$ and the relative position null four-vector $r_j(x)=x-\psi_j (\tau_j(x))$, for each source $j$. Denote the four-velocity and the four-acceleration of the $j$th source at the retarded time $\tau_j(x)$ by $w_j(x)$ and $a_j(x)$, respectively.

With this notation, for each source $j$, the   four-potential  is
\begin{equation}\label{lWhitj}
l_j(x)=\frac{r^{*w_j}_j(x)}{(r_j(x)\cdot w_j(x))^{3/2}}=\frac{2(r_j(x)\cdot w_j(x))w_j(x)-r_j(x)}{(r_j(x)\cdot w_j(x))^{3/2}},
\end{equation}
and  the gravitational potential tensor is
\begin{equation}
    h_{j:\alpha\beta}=2M_jl_{j:\alpha} (x)l_{j:\beta} (x),
\end{equation}
where the notation $l_{j:\alpha}$ means the $\alpha$ component of the four-covector $l_j$. 

Since each tensor $h_j$ is proportional to the mass $M_j$ of the source, we may assume that the gravitational potential tensor of the combined field is
\[  h_{\alpha\beta}(x)= \sum_j 2M_jl_{j:\alpha} (x)l_{j:\beta} (x),
\]
the sum of the gravitational potential tensors of each source. This implies that the {\em metric for the combined field}  is
\begin{equation}\label{metricWcomp}
    g_{\alpha\beta}(x)= \eta_{\alpha\beta}-\sum_j 2M_jl_{j:\alpha} (x)l_{j:\beta} (x).
\end{equation}

To obtain the equation of motion in such a field, we need to compute the inverse metric $g^{\alpha\beta}(x)$ of the metric \eqref{metricWcomp}. This may be done as follows. Rewrite \eqref{metricWcomp} as
\[g_{\alpha\beta}=\eta_{\alpha\mu}(\delta^\mu_\beta-H^\mu_\beta),\]
 where  the operator $H$ of the relativistic gravitational potential of the combined field is
\begin{equation}\label{Adef}
 H^\mu_\beta=\sum_j 2M_jl_j^\mu l_{j:\beta} .   
\end{equation}
This operator is of first order in the masses of the sources.  Direct verification shows that the inverse metric is 
\begin{equation}\label{metrInv}
g^{\alpha\beta}=(I-H)^{-1}\eta^{\alpha\beta}.
 \end{equation}

\section{The equation of motion in a gravitational field of several sources}
$\;\;\;\;\;$
In $GR$ and in differential geometry, one normally uses the arc length parameter $s$ for the evolution parameter. However, this parameter depends on the metric of the field.  Hence, to describe motion in a gravitational field generated by more than one source, the use of the arc length parametrization, which is different for each source, complicates the problem. Indeed, it is not reasonable to add accelerations measured by clocks running at different rates. This is why we use the Minkowskian parametrization, which is independent of the field. Since the acceleration due to each source will be with respect to the same parameter $\tau$, we will be able to add the accelerations. The used of Minkowskian parametrization is also needed in the next section to obtain explicit formulas for the dynamics in the gravitational field.

We assume that the motion of an object in the field is along a geodesic determined by a metric $g_{\mu\nu}(x)$.  From the Euler-Lagrange equations, the geodesic equation is obtained as follows.  First, we calculate the \emph{Christoffel symbols of the first kind} 
\[\Gamma_{\alpha\mu\nu}=\frac{1}{2}(g_{\alpha\mu,\nu}+g_{\alpha\nu,\mu}-g_{\mu \nu,\alpha}).\]
In $GR$ this is not a tensor, since the differentiation of the metric uses partial derivatives and not covariant derivatives. Nevertheless, in our model, which is Lorentz-covariant, partial derivatives of a Lorentz-covariant metric are a tensor, implying that $\Gamma_{\alpha\mu\nu}$ is a tensor of rank $(0,3)$ under the Lorentz transformations on Minkowski space.

Next, the \emph{Christoffel symbols of the second kind}, sometimes simply referred to as the \emph{Christoffel symbols}, are defined by
\begin{equation}\label{GamGtens}
\Gamma^{\lambda}_{\mu\nu}=g^{\lambda\beta}\Gamma_{\beta\mu\nu}.
\end{equation}
This symbol is not a tensor is our model, since the raising of the index is done by the metric $g$ and not by the Minkowski metric. 
The well-known geodesic equation with respect to the arc length parameter $s$ is
\begin{equation}\label{GeodesicS}
 \frac{d^2x^\lambda}{ds^2} =-\Gamma^{\lambda}_{\mu\nu}\frac{dx^\mu}{ds}\frac{dx^\nu}{ds}.   
\end{equation}
As shown in \cite{Gron}, this equation can be rewritten in the form 
\[\frac{d^2x_\lambda}{ds^2} =\frac{1}{2}g_{\mu\nu,\lambda}\frac{dx^\mu}{ds}\frac{dx^\nu}{ds}.\]
This shows that the covariant four-velocity components $u_\lambda=\frac{dx_\lambda}{ds}$ conjugate to cyclic coordinates are conserved.

We now compute the geodesic equation with respect to the Minkowski parameter. Using the chain rule, we have
\[\frac{dx}{ds}=\frac{dx}{d\tau}\frac{d\tau}{ds}, \; \frac{d^2x}{ds^2}=\frac{d^2x}{d\tau^2}\left(\frac{d\tau}{ds}\right)^2+\frac{dx}{d\tau}\frac{d^2\tau}{ds^2} .\]
Using \eqref{GeodesicS}, we obtain
\[\frac{d^2x^\lambda}{d\tau^2}\left(\frac{d\tau}{ds}\right)^2+\frac{dx^\lambda}{d\tau}\frac{d^2\tau}{ds^2}=-\Gamma^{\lambda}_{\mu\nu}\frac{dx^\mu}{d\tau}\frac{dx^\nu}{d\tau}\left(\frac{d\tau}{ds}\right)^2.\]
Multiplying this equation by $\left(\frac{ds}{d\tau}\right)^2$ yields
\[\frac{d^2x^\lambda}{d\tau^2}+\frac{dx^\lambda}{d\tau}\frac{d^2\tau}{ds^2}\left(\frac{ds}{d\tau}\right)^2=-\Gamma^{\lambda}_{\mu\nu}\frac{dx^\mu}{d\tau}\frac{dx^\nu}{d\tau}.\]
Contracting this equation with $\dot{x}_\lambda= \frac{dx_\lambda}{d\tau}$ and using  $\dot{x}\cdot \dot{x}=1$ and $\ddot{x}\cdot \dot{x}=0$, we obtain
\[\frac{d^2\tau}{ds^2}\left(\frac{ds}{d\tau}\right)^2=-\Gamma^{\lambda}_{\mu\nu}\frac{dx^\mu}{d\tau}\frac{dx^\nu}{d\tau}\frac{dx_\lambda}{d\tau}.\]
Finally, the geodesic equation (see also  \cite{Bel}) with respect to the Minkowskian parameter is
\begin{equation}\label{GeodesicMdot0}
 \ddot{x}^\lambda =-\Gamma^{\lambda}_{\mu\nu}\dot{x}^\mu\dot{x}^\nu+\Gamma^{\alpha}_{\mu\nu}\dot{x}^\mu\dot{x}^\nu\dot{x}_\alpha\dot{x}^\lambda.   
\end{equation}
The first term, which we denote by $\ddot{x}_{(q)}$, is quadratic in the four-velocity of the moving object. Then
\begin{equation}\label{GeodesicMdot}
 \ddot{x}=\ddot{x}_{(q)}-(\ddot{x}_{(q)}\cdot \dot{x})\dot{x},\;\;\; \ddot{x}_{(q)}^\lambda=-\Gamma^{\lambda}_{\mu\nu}\dot{x}^\mu\dot{x}^\nu.   
\end{equation}
The second term ensures that acceleration is perpendicular to the four-velocity in Minkowski space. This equation shows that the dynamics in an arbitrary gravitational field with respect to the Minkowskian parameter is determined by the Christoffel symbols $\Gamma^{\lambda}_{\mu\nu}$.

To find the acceleration in the combined field, we  use formula \eqref{GeodesicMdot}, which depends on the Christoffel symbols $\Gamma^{\alpha}_{\mu\nu}$. We define first the symbol $\Gamma_{j:\alpha\mu\nu}$ for each source $j$. Since the derivatives of metric \eqref{metricWcomp} are linear in the sources, it follows that the Christoffel symbols of the first kind for the combined field are the {\em sum} of the corresponding symbols of each source. Thus, $\Gamma_{\alpha\mu\nu}$ of the combined field is 
\[    \Gamma_{\alpha \mu\nu}=  \sum_j \Gamma_{j:\alpha\mu\nu}\]
and is a tensor of rank $(0,3).$

We define the {\em first-order tensor} $G^{\alpha}_{\mu\nu}$ of the field by raising the first index in the above tensor using the Minkowski metric. Thus,
\begin{equation}\label{Gdef}
 G^{\alpha}_{\mu\nu}=-\eta^{\alpha\beta} \Gamma_{\beta\mu\nu}=\frac{1}{2}\eta^{\alpha\beta}(g_{\mu \nu,\beta}-g_{\beta\mu,\nu}-g_{\beta\nu,\mu})   
\end{equation}
is a rank $(1,2)$ tensor with respect to the Lorentz transformations.
For the combined field, the first-order tensor is
\[   G^{\alpha}_{\mu\nu}=\sum_j G^{\alpha}_{j:\mu\nu}\,,\]  
the sum of the first-order tensors of each source.

Substituting \eqref{metrInv} and \eqref{Gdef} into \eqref{GamGtens}, the Christoffel symbols $\Gamma^{\alpha}_{\mu\nu}$ of the combined field are 
 \begin{equation}\label{GamComb} -\Gamma^{\alpha}_{\mu\nu}=((I-H)^{-1})^\alpha_\beta G^{\beta}_{\mu\nu}\,,
 \end{equation}
 where  $H$ is defined by \eqref{Adef}. This formula explicitly defines the decomposition of the Christoffel symbols into a power series in the mass of the sources, where the $k$th order term is
 \[G^{(k)\alpha}_{\mu\nu}=(H^{k-1})^\alpha_\beta G^{\beta}_{\mu\nu}\,\] and
  \[-\Gamma^{\alpha}_{\mu\nu}=\sum_{k=1}^\infty  G^{(k)\alpha}_{\mu\nu}.\]
The {\em first-order quadratic acceleration} is defined by
\begin{equation}\label{accelFirstOrdComb}
\ddot{x}^{(1)\lambda}_{(q)}=G^{\lambda}_{\mu\nu}\dot{x}^\mu\dot{x}^\nu.    
\end{equation}
Then, the quadratic acceleration is 
 \begin{equation}\label{AccelCom}
   \ddot{x}_{(q)}=(I-H)^{-1} \ddot{x}^{(1)}_{(q)}\,,
\end{equation}
and \eqref{GeodesicMdot} defines the acceleration $\ddot{x}$ of an object moving in the combined field.

This acceleration differs from the Newtonian acceleration because it includes the dependence on the motion of the source and on the velocity of the moving object.

We point out several nice properties of our model. First, equation  \eqref{AccelCom} implies that if at some spacetime point $x$, we have $\ddot{x}^{(1)}_{(q)}=0$, then $\ddot{x}=0$ at $x$ as well. Next, since $G^{\alpha}_{\mu\nu}$ is linear in the sources, the first-order quadratic acceleration of the combined field is
 \begin{equation}\label{AcceldecompFirst}
  \ddot{x}^{(1)}_{(q)}=\sum_j \ddot{x}^{(1)}_{j:(q)},   
 \end{equation}
the sum of the first-order quadratic accelerations due to each source.
One easily notices that the acceleration in the combined field is  \emph{not} linear in the sources, since $H$ depends on all of the sources. Nevertheless, this acceleration can be computed from the first-order quadratic acceleration, using \eqref{AccelCom}. This is the reason we call our model {\em quasi-linear}. Finally,
all of our derivations may be extended to an arbitrary gravitational field, where the sources are continuously distributed dynamic masses. To do this, one replaces the summation by $j$ with integration over the backward light cone with vertex at $x$ of spacetime positions $x'$ of the sources and  replace their mass with the mass density at this position.

 \section{The near and far components of a  gravitational field}
$\;\;\;\;\;$
Since the acceleration in a combined field is a function of the first-order tensor, which, in turn, is the sum of the first-order tensors of each source, we derive here the explicit form of the first-order tensor of a gravitational field of a single source of mass $M$. For this,  
 we first compute the Christoffel symbols of the first kind.  Using $g_{\mu\nu,\alpha}=-2M(l_\mu l_{\nu,\alpha}+l_\nu l_{\mu,\alpha})\, ,$ we obtain
\[- \Gamma_{\alpha\mu\nu}=M(l_\mu l_{\alpha,\nu} +l_\alpha l_{\mu,\nu}+l_\nu l_{\alpha,\mu}+l_\alpha l_{\nu,\mu}-l_\nu l_{\mu,\alpha}-l_\mu l_{\nu,\alpha}). \]
This symbol is symmetric in $\mu$ and $\nu$. Thus,  \begin{equation}\label{Gamma1wl}
    -\Gamma_{\alpha\mu\nu}=M(l_\mu l_{\alpha,\nu} -l_\mu l_{\nu,\alpha}+l_\alpha l_{\mu,\nu})^s,
\end{equation}
where $s$ denotes symmetrization with respect to $\mu$ and $\nu$. To simplify the symmetrization notation, we introduce, for any 2 covectors $x$ and $y$, a symmetrized  rank $(0,2)$ tensor $x\vee y$, defined by
\begin{equation}\label{vee}
(x\vee y)_{\mu\nu}=x_\mu y_\nu+x_\nu y_\mu.
\end{equation}
Contracting twice with a four-vector $u$ yields
\[(x\vee y)_{\mu\nu}u^\mu u^\nu=2(x\cdot u)(y\cdot u).\]

Since the four-vector-valued function $l(x)$, defined by  \eqref{lWhit0}, depends on the retarded position $r(x)$ and the source's retarded velocity $w(x)$, we need explicit formulas for their partial derivatives with respect to $x^\mu$. 
From \cite{ana}, page 60, formulas (3.86) and (3.88), and \cite{Jackson}, we have
\begin{equation}\label{rnulower}
   \tau(x)_{,\mu}=\frac{r_\mu}{r\cdot w}\;\;\;,\;\;\;r_{\nu,\mu}=\eta_{\nu\mu}-\frac{w_\nu r_\mu}{r\cdot w}. 
\end{equation}
For the derivatives of the four-velocity of the source, we have
\begin{equation}\label{Der4Vel}
w_{\nu,\mu}=\frac{a_\nu r_\mu}{r\cdot w},
\end{equation}
where $a_\nu(\tau(x))$ is the source four-acceleration covector at the retarded time, and
\begin{equation}\label{rDotw2}
(r\cdot  w)_{,\mu}=w_\mu+\frac{r_\mu((a\cdot  r)-1)}{r\cdot w }.
\end{equation}

 From (\ref{lWhit0}), 
 \[l_{\nu,\mu}=\frac{2(r\cdot w)_{,\mu}w_\nu +2(r\cdot w)w_{\nu,\mu}-r_{\nu,\mu}}{ (r\cdot w)^{3/2}}-\frac{3}{2}\frac{(2(r\cdot w)w_\nu-r_\nu)(r\cdot w)_{,\mu} }{(r\cdot w)^{5/2}},
 \]
 or
  \[l_{\nu,\mu}=\frac{2(r\cdot w)w_{\nu,\mu}-r_{\nu,\mu}}{ (r\cdot w)^{3/2}}+\frac{(-(r\cdot w)w_\nu+1.5r_\nu)(r\cdot w)_{,\mu} }{(r\cdot w)^{5/2}}.
 \]

Using  \eqref{rnulower}, \eqref{Der4Vel}  and  \eqref{rDotw2}, we split this expression into two components. The first component, called the \emph{near component}, contains all terms without $a$, and the second component, which we call the \emph{far component}, contains all terms involving $a$. These components are 
 \begin{equation}\label{nearcomder}
  l^{(n)}_{\nu,\mu}= \sqrt{\frac{1}{(r \cdot w)^3}}\left( -w_\nu w_\mu-\eta_{\mu\nu}+\frac{2 w_\nu r_\mu}{r\cdot w}+\frac{3}{2}\frac{r_\nu w_\mu }{r\cdot w}-\frac{3}{2}\frac{ r_\nu r_\mu}{(r\cdot w)^2}\right)
 \end{equation}
and
\begin{equation}\label{farcomder}
  l^{(f)}_{\nu,\mu}= \sqrt{\frac{1}{(r \cdot w)^3}}\left(2a_\nu r_\mu-\frac{ w_\nu r_\mu (a\cdot  r)}{r\cdot w }+\frac{3}{2}\frac{r_\nu r_\mu (a\cdot  r)}{(r\cdot w)^2 }\right).  
 \end{equation}
 This implies that
\[l_\alpha l^{(n)}_{\nu,\mu}= \frac{2(r\cdot w)w_\alpha-r_\alpha}{(r \cdot w)^3}\left( -w_\nu w_\mu-\eta_{\mu\nu}+\frac{2 w_\nu r_\mu}{r\cdot w}+\frac{3}{2}\frac{r_\nu w_\mu }{r\cdot w}-\frac{3}{2}\frac{ r_\nu r_\mu}{(r\cdot w)^2}\right) \]
and
\[l_\alpha l^{(f)}_{\nu,\mu}= \frac{2(r\cdot w)w_\alpha-r_\alpha}{(r \cdot w)^3}\left( 2a_\nu r_\mu-\frac{ w_\nu r_\mu (a\cdot  r)}{r\cdot w }+\frac{3}{2}\frac{r_\nu r_\mu (a\cdot  r)}{(r\cdot w)^2 }\right). \] 

To obtain the Christoffel symbols of the first kind for our field, we substitute the previous expressions in \eqref{Gamma1wl}. To simplify the substitution, note that the combination $l_\mu(l_{\alpha,\nu}-l_{\nu,\alpha})$ of the first two terms in \eqref{Gamma1wl} is antisymmetric in $\alpha,\nu$, and so their near component is
\[l_\mu(l^{(n)}_{\alpha,\nu}-l^{(n)}_{\nu,\alpha})=\frac{1}{2}\frac{2(r\cdot w)w_\mu-r_\mu}{(r\cdot w)^4} (r_\nu w_\alpha
-w_\nu r_\alpha ),\] 
and the far component is
\[l_\mu(l^{(f)}_{\alpha,\nu}-l^{(f)}_{\nu,\alpha})= \frac{2(r\cdot w)w_\mu-r_\mu}{(r\cdot w)^3} \left(2(r_\nu a_\alpha
-a_\nu r_\alpha)-\frac{a\cdot r}{r\cdot w}(r_\nu w_\alpha
-w_\nu r_\alpha) \right ).\] 

Now add the last term of \eqref{Gamma1wl} and symmetrize with respect to $\mu$ and $\nu$ to obtain $\Gamma_{\alpha \mu\nu}$. By raising the index $\alpha$, we obtain first-order tensor $G^\alpha _{\mu\nu}$, defined by \eqref{Gdef}. For the near component $G^{(n)}$, we use notation \eqref{vee} and separate the directions $r^\alpha$ and  $w^\alpha$.  This gives
\[ G^{(n)} =M\left(\frac{2\eta}{(r\cdot w)^3} -\frac{3 w\vee r}{(r\cdot w)^4}+\frac{3}{2}\frac{ r\vee r}{(r\cdot w)^5}\right)r  \]
\begin{equation}\label{tensNear}
    +M\left(\frac{-4\eta-2w\vee w}{(r\cdot w)^2} +\frac{8 w\vee r }{(r\cdot w)^3}-\frac{7 r\vee r}{2(r\cdot w)^4}\right)w\,. 
\end{equation}
If we separate the near component into the $r^{*w}$ direction, defined by \eqref{defConj}, and $w$, we obtain 
\[ G^{(n)} =-M\left(\frac{2\eta}{(r\cdot w)^3} -\frac{3 w\vee r}{(r\cdot w)^4}+\frac{3}{2}\frac{ r\vee r}{(r\cdot w)^5}\right)(r^{*w}) + \]
\begin{equation}\label{tensNear2}
    M\left(\frac{-4w\vee w}{(r\cdot w)^2} +\frac{2 w\vee r }{(r\cdot w)^3}-\frac {r\vee r}{(r\cdot w)^4}\right)w\,. 
\end{equation}

Denote by $u=\dot{x}$ the velocity of the object. Then the first-order quadratic acceleration $\ddot{x}^{(1)(n)}_{(q)}=G^{(n)\alpha}_{\mu\nu}u^\mu u^\nu$ caused by the near field is, from \eqref{tensNear},
\[\ddot{x}^{(1)(n)}_{(q)}=M\left(\frac{2}{(r\cdot w)^3} -\frac{6 (w\cdot u )(r\cdot u)}{(r\cdot w)^4}+\frac{3 (r\cdot u)^2}{(r\cdot w)^5}\right)r \]
\begin{equation}\label{an}
  +M\left(\frac{-4(1+(w\cdot u)^2)}{(r\cdot w)^2} +\frac{16 (w\cdot u )(r\cdot u)}{(r\cdot w)^3}-\frac{7 (r\cdot u)^2}{(r\cdot w)^4}\right)w.  
\end{equation}

Let us check the Newtonian limit of this formula. In this limit, $w=u=(1,0,0,0)$ and $r=(R,\mathbf{R})=R(1,\mathbf{n})$, which implies that $r\cdot w= r\cdot u=R, w\cdot u=1$. Hence, the spatial part of the first-order acceleration is approximately  $ -\frac{M}{R^2}\mathbf{n}$. Using \eqref{mass} and the fact that for small velocities, $d\tau=c dt$, the usual acceleration is
\[\frac{d^2\mathbf{x}}{dt^2}=-\frac{Gm}{R^2}\mathbf{n},\]
as in Newtonian gravity. 
The purely relativistic term of $w$ expresses the transfer of the momentum of the source to the field and is similar to field dragging. For example, the gravitational field of a star rotating around the black hole of a galaxy will cause an additional rotational velocity to other stars in the galaxy. Note that for large distances $R$ from the source, all terms in the near component fall off like $1/R^2$.

Repeating the same procedure for the far component $G^{(f)}$, separating the directions $r^\alpha$,  $w^\alpha$ and $a^\alpha$, and simplifying yields
\[ G^{(f)} =M\left(\frac{-4a\vee w}{(r\cdot w)^2} +\frac{2 (r\cdot a)w\vee w}{(r\cdot w)^3}-\frac{3}{2}\frac{(r\cdot a) r\vee r}{(r\cdot w)^5}\right)r  \]
\[  +M\left(\frac{4(a\vee r)}{(r\cdot w)^2} -\frac{4(r\cdot a) w\vee r }{(r\cdot w)^3}+\frac{4(r\cdot a)r\vee r}{(r\cdot w)^4}\right)w \]
\begin{equation}\label{tensFar}
  +M\left(\frac{4(w\vee r)}{(r\cdot w)^2} -\frac{2r\vee r}{(r\cdot w)^3}\right)a . 
\end{equation}

The first-order quadratic acceleration  caused by the far  field is 
\[ \ddot{x}^{(1)(f)}_{(q)} =M\left(\frac{-8(a\cdot u)(w\cdot u)}{(r\cdot w)^2} +\frac{4 (r\cdot a)(w\cdot u)^2}{(r\cdot w)^3}-\frac{3(r\cdot a) (r\cdot u)^2}{(r\cdot w)^5}\right)r  \]
\[  +8M\left(\frac{(a\cdot u)(r\cdot u)}{(r\cdot w)^2} -\frac{(r\cdot a) (w\cdot u)(r\cdot u) }{(r\cdot w)^3}+\frac{(r\cdot a)(r\cdot u)^2}{(r\cdot w)^4}\right)w \]
\begin{equation}\label{accelFar}
  +4M\left(\frac{2(w\cdot u)(r\cdot u) }{(r\cdot w)^2} -\frac{(r\cdot u)^2}{(r\cdot w)^3}\right)a . 
\end{equation}
Note that for large distances $R$ from the source, all terms in the far component fall off like $1/R$.

The first-order quadratic acceleration for a single-source gravitational field, defined by \eqref{accelFirstOrdComb}, becomes
\begin{equation}\label{accelFirst OrdFinal}
\ddot{x}^{(1)}_{(q)}= \ddot{x}^{(1)(n)}_{(q)}+\ddot{x}^{(1)(f)}_{(q)},  
\end{equation}
where $\ddot{x}^{(1)(n)}_{(q)}$ and $\ddot{x}^{(1)(f)}_{(q)}$ are defined by \eqref{an} and \eqref{accelFar}.

In addition to the first-order gravitational field,  a single-source gravitational field also has a non-linear part. Since $l$ is a null covector, for such a field, the operator $H$ of the gravitational potential, defined by \eqref{Adef}, satisfies $H^2=0$. From \eqref{GamComb}, it follows that the non-linear part of the Christoffel symbols is only of order two in $M$.

For a combined field, the first-order tensor is the sum of the first-order tensors of the near field, defined by \eqref{tensNear}, and the corresponding tensor \eqref{tensFar} of the far field. The combined tensor can be also decomposed into near and far components.

For a static source $M_j$ resting at $\mathbf{x}_j$, the acceleration of an object at $x$, which is temporally at rest, we have $\dot{x}=(1,0,0,0)$, and the first-order acceleration is
\[\ddot{x}_j^{(1)}(\mathbf{x})=\left(0,-M_j\frac{\mathbf{x}-\mathbf{x}_j}{|\mathbf{x}-\mathbf{x}_j|^3}\right)=(0,\mathbf{a}_j (\mathbf{x})),\]  where $ c^2\mathbf{a}_j (\mathbf{x})$ is the Newtonian acceleration caused by this source. By \eqref{AcceldecompFirst}, the first-order acceleration caused by the combined field of resting sources is 
\[\ddot{x}^{(1)}=\sum_j \ddot{x}_j^{(1)}=\sum_j(0, \mathbf{a}_j (\mathbf{x}))=(0,\mathbf{a} (\mathbf{x}))\,,\]
where the last equality follows from the fact that the acceleration of the combined field is the sum of the accelerations caused by each source. Thus, for a static gravitational field, the linearity of the first-order acceleration is equivalent to the superposition principle in Newtonian gravity.

\section{Comparison between electromagnetic and gravitational near and far fields.}
$\;\;\;\;\;$
To compare the motion of an object in a single-source gravitational field with the motion of a charge in a single-source electromagnetic field, we will consider the motion  of a charge $q$ with mass $m$  in an electromagnetic field generated by a moving charge $Q$. Such a field can be described by the Li\'{e}nard-Wiechert four-potential 
\begin{equation}\label{Form 1 pote}
 A_\mu (x)=\frac{Q}{4\pi\epsilon_0}\frac{w_\mu(\tau(x))}{r(x)\cdot  w(\tau(x))}.
\end{equation}
Using our notation, the acceleration in such a field with respect to the parameter $\tau$ is given by
\[\ddot{x}^\lambda=k\eta^{\lambda\mu}(A_{\nu,\mu}-A_{\mu,\nu})\dot{x}^\nu,\]
where $k=q/mc^2$. 
From the formulae \eqref{rnulower}-\eqref{rDotw2} above, we have
\[ A_{\nu,\mu}=\frac{Q}{4\pi\epsilon_0}\left[\frac{a_\nu r_\mu}{(r\cdot w)^2} -\frac{w_\nu w_\mu}{(r\cdot w)^2}+\frac{w_\nu r_\mu}{(r\cdot w)^3} -\frac{w_\nu r_\mu(a\cdot  r)}{(r\cdot w)^3}  \right].\]

Also here, the field splits into a near field, consisting of the terms not containing the acceleration $a$ of the source, and a far field, consisting of the terms containing $a$. All of the terms of the near field fall of like $1/R^2$ at large distances $R$, while the terms of the far field fall off like $1/R$ at large distances $R$.
The acceleration caused by the near field is 
\begin{equation}\label{accelEMnear}
  \ddot{x}^{(n)}=\mathcal{E}\left(\frac{w\cdot u }{(r\cdot w)^3}r-\frac{r\cdot u }{(r\cdot w)^3}w \right), \;\; \mathcal{E}= \frac{g}{mc^2}\frac{Q}{4\pi\epsilon_0}
\end{equation}
where $u=\dot{x}$ is the four-velocity of the test charge. 

Let us compare this acceleration to the corresponding first-order quadratic acceleration \eqref{an} in a single-source gravitation field. 
The Newtonian limits are very similar. In both cases, this acceleration is a linear combination of $r$ and $w$. Writing the the four-velocity $u$ as $u=\gamma(\beta)(1,\boldsymbol{\beta})$, in electromagnetism we can decompose this acceleration into two parts. The part connected to the electric field is independent of $\boldsymbol{\beta}$ (but depends on $\gamma(\beta)$), and the part connected to the magnetic field is linear in $\boldsymbol{\beta}$. For gravity, a similar procedure reveals that there is a part of the acceleration  which is independent of $\boldsymbol{\beta}$. This part may be called the gravitoelectric field. There are two more parts, one linear in $\boldsymbol{\beta}$ and one quadratic in $\boldsymbol{\beta}$. 
It is not obvious that the linear in $\boldsymbol{\beta}$ part can be considered as gravitomagnetic, since the 3D acceleration due to the magnetic field is perpendicular to $\boldsymbol{\beta}$, which may not hold for the linear part of the gravitation near field.

The far electromagnetic field of an accelerated source generates electromagnetic waves. To compare this field with the corresponding gravitational field, decompose its acceleration into $r,w,a$ components. This yields
\begin{equation}\label{EMfar}
\ddot{x}^{(f)}=\mathcal{E}\left(\frac{(r\cdot w)(a\cdot u)-(a\cdot r)(w\cdot u)}{(r\cdot w)^3}r+\frac{(a\cdot r)(r\cdot u)}{(r\cdot w)^3}w-\frac{(r\cdot u)}{(r\cdot w)^2}a\right). \end{equation}
Comparing this formula with \eqref{accelFar}, the corresponding formula for the gravitation field, we observe that in both cases the far acceleration is a combination of vectors $r,w$ and $a$ and depends linearly on $a$. However, their dependence on the four-velocity $u$ and the 3D velocity of the object is very different. 

Let us compare the far field generated by a binary star rotating in a circular orbit far from the observer with a far field from a pair of rotating charges on similar orbits.

We assume that the center of mass of the binary star is at rest at the space origin in $K$ and the $x,y$ plane is the plane of the binary. Since the field of such binary with circular orbits is symmetric with respect to rotation about the $z$-axis, to define the far field of binary, it is enough to define its acceleration on an object at  points $x=(ct,R_0\sin{\theta},0,R_0\cos{\theta})$, for arbitrary $t,R_0,\theta$. We assume that $R_0$ is much larger than the size of the binary, implying that the relative position of each star at the retarded time is
\[r=(R_0,R_0\sin{\theta},0,R_0\cos{\theta}).\] 

Let $R_1$ and $R_2$, respectivelt, denote the radii of the trajectories of two stars of masses $m_1$ and $m_2$. From the definition of the center of mass, we have 
\[R_1=\frac{m_2 R}{M},\;\;R_2=\frac{m_1 R}{M}, \;\;\mbox{for} \;\; R=R_1+R_2, M=m_1+m_2.\]
We ignore terms of order $\frac{R_1^2\Omega^2}{c^2}$ in $\gamma$ terms and assume that $\gamma=1.$ By a time shift, if necessary, we may assume, without loss of generality, that the retarded velocity of the first star is
\[w_1=e_0 +R_1\widehat{w},\;\;e_0=(1,0,0,0),\;\; \widehat{w}=\frac{\Omega}{c}(0, -\sin{\Omega t},\cos{\Omega t},0). \]
Thus, its retarded acceleration is  
\[a_1=R_1 \hat{a},\;\; \hat{a}=\frac{\Omega^2}{c^2}(0, -\cos{\Omega t},-\sin{\Omega t},0) .\]
Similarly, the retarded velocity and acceleration of the  second star are
\[w_2=e_0 -R_2\widehat{w}, \;\;
a_2=-R_2 \hat{a}. \]

We assume that the four-velocity of the object is small, so that $u=(1,0,0,0)$. To apply formula \eqref{accelFar} for the
 first-order quadratic acceleration for a single-source gravitational field, we need the following expressions:
 \[r\cdot u=R_0,\;\; w_j\cdot u=1,\;\; a_j\cdot u=0,\]
 for $j=1,2$,
 \[r\cdot w_1=R_0(1+R_1\widehat{rw}),\;
 r\cdot w_2=R_0(1-R_2\widehat{rw}),\;\widehat{rw}=\frac{\Omega}{c}\sin{\Omega t}\sin{\theta}\]
and
 \[r\cdot a_1=R_0 R_1\widehat{ra},\;\;
 r\cdot a_2=-R_0 R_2\widehat{ra},\;\; \widehat{ra}=\frac{\Omega^2}{c^2}\cos{\Omega t}\sin{\theta}.\]
 Note that 
 \[m_1 a_1=-m_2 a_2 =-\frac{m_1 m_2 R}{M }\hat{a}\] 
 and
 \[m_1(r\cdot a_1)=-m_2(r\cdot a_2)=\frac{m_1 m_2R_0 R}{M }\widehat{ra}.\]

Formula \eqref{accelFar} has eight terms. The first and the fourth terms are zero, since $a\cdot u=0.$ To obtain the formula for the acceleration caused by the combined field, we will add the acceleration caused by each term in this formula for both stars. To do this, we assume that the unit-free velocity of the stars is small and use the approximation $\frac{1}{(1+\epsilon)^n}\approx 1-n\epsilon$ for $\frac{1}{(r\cdot w_j)^n}$. This yields
 \begin{equation}\label{rwndif}
   \frac{k_1R^n_0}{(r\cdot w_1)^n} + \frac{k_2R^n_0}{(r\cdot w_2)^n}= k_1+k_2-n(k_1R_1-k_2R_2)\widehat{rw},
 \end{equation}
 and
 \[\frac{k_1R_0^n w_1}{(r\cdot w_1)^n} + \frac{k_2 R_0^n w_2}{(r\cdot w_2)^n}= (k_1+k_2-n(k_1R_1+k_2R_2)\widehat{rw})e_0\]
\begin{equation}\label{rwwdif}
   +(k_1R_1-k_2R_2-n(k_1R_1^2+k_2R_2^2)\widehat{rw})\widehat{w}.
 \end{equation}
 
 To define the  $r$  component of the combined field in \eqref{accelFar}, we use \eqref{rwndif} with $k_1=-k_2= 4Gm_1(r\cdot a_1)/c^2$ and $n=3$ for the second term, and $k_1=-k_2= -3Gm_1(r\cdot a_1)/c^2$ and $n=5$ for the third term. This yields
 \[\ddot{x}^{(f)}_{(r)}=\frac{3 Gm_1 m_2R^2\Omega^3}{2Mc^5 R_0}\sin{2\Omega t}\sin^2{\theta}(1,\sin{\theta},0,\cos{\theta}).\]
A similar calculation for the $a$ component, applying \eqref{rwndif} with $k_1=-k_2= 8Gm_1 a_1/c^2$ and $n=2$ for the first term in this component, and $k_1=-k_2= -4Gm_1 a_1/c^2$ and $n=3$ for the third term, yields
  \[\ddot{x}^{(f)}_{(a)}=\frac{4 Gm_1 m_2R^2\Omega^3}{Mc^5 R_0}\sin{\theta}\sin{\Omega t}(0,\cos{\Omega t},\sin{\Omega t},0).\]
 For the calculation of the $w$ component, we use \eqref{rwwdif} with 
 with $k_1=-k_2= -8Gm_1(r\cdot a_1)/c^2$ and $n=3$ for the first term in this component, and $k_1=-k_2= 8Gm_1(r\cdot a_1)/c^2$ and $n=4$ for the second term. This yields
\[\ddot{x}^{(f)}_{(w)}=\frac{8 Gm_1 m_2R^2\Omega^3}{Mc^5 R_0}((R_1-R_2)(\Omega/c)\sin^2{\theta}\sin{2\Omega t}\,e_0\]\[-(R_1^2-R_2^2)(\Omega^2/c^2) \sin^2{\theta}\sin{2\Omega t}(0,-\sin{\Omega t},\cos{\Omega t},0))\]
Adding these three components define the first-order quadratic acceleration caused by the far field of a binary star on mirrors of a gravitational wave detector. Since $H\ll I$ from \eqref{AccelCom} it follows that this is also approximately the quadratic acceleration. The spacial part of the full acceleration, defined by \eqref{GeodesicMdot}  in this case is the same a the quadratic one, that we obtained. By integrating twice, we obtain the displacement between the mirrors.  

 We calculate now similar acceleration terms for the far electromagnetic field due to a pair of equal negative charges rotating on a circle of radius $R$ around a positive charge. Formula \eqref{EMfar} yields
  \[\ddot{x}^{(f)}_{(r)}=\frac{3 \mathcal{E}R^2\Omega^3}{c^3 R_0}\sin{2\Omega t}\sin^2{\theta}(1,\sin{\theta},0,\cos{\theta}),\]
  \[\ddot{x}^{(f)}_{(a)}=\frac{4 \mathcal{E}R^2\Omega^3}{c^3 R_0}\sin{\theta}\sin{\Omega t}(0,\cos{\Omega t},\sin{\Omega t},0)\]
  and
  \[\ddot{x}^{(f)}_{(w)}=\frac{2 \mathcal{E}R^2\Omega^3}{c^3 R_0}(R(\Omega/c)\sin^2{\theta}\sin{2\Omega t}\,e_0\]
  \[+ 2R(\Omega/c) \sin{\theta}\cos{\Omega t}(0,-\sin{\Omega t},\cos{\Omega t},0)).\]

  We observe that the far field of a binary star is similar to the far field of an electromagnetic quadruple radiation wave. Two components are similar, but the $w$ component is different.

\section{An Algorithm for calculating motion in a combined gravitational field}
$\;\;\;\;\;$
The following simple and straightforward algorithm calculates the worldline of an object moving freely in a field  generated by a collection of spherically symmetric masses $M_j$, for a finite set of indices $\{j\}$, moving along worldlines $\psi_j(\tau)$. The object  at some initial time $\tau_0$ is at spacetime point $x$, and its four velocity is $\dot{x}$. To know the position and the four-velocity of the object at time $\tau_0+\Delta \tau$, for small $\Delta \tau$, we need to find the four-acceleration caused by the gravitational field.

This can be done by the following steps.
\begin{enumerate}
    \item For each source $j$, solve the equation $(x-\psi_j(\tau))^2=0$ and denote the solution by $\tau=\tau_j(x)$.
    \item For each source $j$, denote the relative position by $r_j=x-\psi_j(\tau_j(x))$, the retarded four-velocity by $w_j(x)=\frac{d\psi_j(\tau)}{d\tau}|_{\tau=\tau_j(x)}$, and the retarded four-acceleration by $a_j(x)=\frac{d^2\psi_j(\tau)}{d\tau^2}|_{\tau=\tau_j(x)}$.
        \item For each source $j$, use \eqref{accelFirst OrdFinal} to obtain the first-order quadratic acceleration $\ddot{x}_{j:(q)}^{(1)}$ caused by the gravitational field of the source $j$.
        \item Use \eqref{AcceldecompFirst} to obtain the first-order quadratic acceleration of the object in the combined field.
    \item Use \eqref{Adef} and \eqref{lWhit0} to calculate the operator $H$.
    \item Use \eqref{AccelCom} and \eqref{GeodesicMdot} to obtain the acceleration of the object in the combined field.
\end{enumerate}

By the standard procedure, set $x(\tau_0+\Delta \tau)=x(\tau_0)+\dot{x}(\tau_0)\Delta \tau$ and $\dot{x}(\tau_0+\Delta \tau)=\dot{x}(\tau_0)+\ddot{x}(\tau_0)\Delta \tau$. In this way, one can compute the worldline of the test object in the field.

\section{Summary and Discussion}
$\;\;\;\;\;$
In this paper, we have described the geometry of a gravitational field generated by a collection of moving, spherically symmetric sources and derived the equation of motion for objects in such a field. To be able to combine the gravitational fields from different sources, we work  on a Minkowski background and need a metric for a single-source gravitational field for which the deviation from the Minkowski metric is proportional to the mass of the source. 
 
 In Section 2, we introduced a new metric for a single, spherically symmetric source which is Lorentz covariant, passes all $GR$ tests, and describes properly behavior near black holes. Its deviation from the Minkowski metric is proportional to the mass of the source.
 In Section 3, we used this metric to describe the geometry of a gravitational field generated by a collection of moving, spherically symmetric sources.  By introducing the operator $H$ of the relativistic gravitational potential, which is linear in the masses of the sources, we derived the inverse of this metric. 
 
 In Section 4, the motion of an object is described by a worldline in a flat background spacetime, parameterized by the Minkowskian parameter. We introduced the first-order tensor $G^{\alpha}_{\mu\nu}$ \eqref{Gdef} of the field  by raising the first index of the Christoffel symbols of the first kind, which is a tensor with respect to the Lorentz transformations. This tensor is linear in the sources of the field and is used to find the first-order (in the masses of the sources)  acceleration. Thus, in our model, both the metric and the first-order acceleration are linear in the sources. The full acceleration is not linear, but may nevertheless be computed directly by applying the operator $(I-H)^{-1}$ to the first-order acceleration. 
 
 In Section 5, for a single-source gravitational field, using known formulas for the partial derivatives of the retarded position and the retarded velocities, we obtained explicit formulas for the near component \eqref{tensNear},  falling off like $1/r^2$, and the far component \eqref{tensFar}, falling off like $1/r$, of the  first-order tensor  and the first-order acceleration \eqref{accelFirst OrdFinal} in such a field. This acceleration is defined by the relative position, four-velocity and four-acceleration of the source at the retarded time and the four-velocity of the the object. We have shown that for a gravitational field generated by several static, spherically symmetric sources, the linearity of the first-order acceleration extends the superposition principle of Newtonian gravity. 
     
 In Section 6, we compared the near and far electromagnetic and gravitational fields. We calculated the acceleration caused by the far gravitational field of a binary star and compared it with quadruple radiation.
In Section 7, we outlined an algorithm for computing the relativistic motion of objects in a combined gravitational field. This is an exact and relatively simple model for relativistic gravitational dynamics. It is not restricted to weak fields, nor to slow-moving sources.  

The existence of a far gravitational field  is important for understanding gravitational waves and the formation of galaxies. Since the source of gravitational waves is the field of collapsing binaries or black holes, which are far from the detector, only their far field will have influence on the detector. As we have seen, in such a case, the assumption that the acceleration caused by the binary is the sum of the accelerations caused by each star or black hole, is supported by the data from the observed $GW$ events.  

Generally, the center of a galaxy is a supermassive black hole. Stars rotating around this black hole form the bulge of the galaxy. In most cases, the mass of the bulge is much larger than the mass of the central black hole. Thus, the stars in the galaxy outside the bulge are influenced mainly by the gravitational field generated by the rotating stars in the bulge. Due to the large distances in the galaxy, also here we expect to observe the effect of the far field and other relativistic corrections of rotating stars. Usual approximation schemes, used in $GR$, cannot be applied
to this problem.

\section{Acknowledgements}
$\;\;\;\;\;$
The author wishes to thank Sergei Kopeikin for important discussions and comments,  Menachem Steiner and Tzvi Scarr  for editing the manuscript.



\begin{thebibliography}{95}
\bibitem{MTW}  C. Misner,  K. Thorne and J. Wheeler,  \emph{Gravitation} (Freeman, San Francisco, 1973)
\bibitem{Adler}R. Adler, M. Bazin and  M. Schiffer  \emph{Introduction to General Relativity}  (McGraw Hill Inc., 1975)
\bibitem{Gron} {\O}. Gr{\o}n, \textit{Introduction to Einstein’s Theory of Relativity.}  (Springer, New York, 2000)
\bibitem{HEL} Hobson, M. P., Efstathiou, G. and Lasenby, A. N. \emph{General Relativity, An Introduction for Physicists}. Cambridge University Press, 2007.
\bibitem{Kop}  S. Kopeikin, M. Efroimsky and G. Kaplan,  \emph{Relativistic Celestial Mechanics of the Solar System} (Wiley‐VCH Verlag GmbH \& Co. KGaA, 2011) 
 \bibitem{Bel} L. Bel, T. Damour, N. Deruelle, J. Ibanez and J. Martin, Poincar\'{e}-Invariant Gravitational Field and Equations of Motion of Two Pointlike Objects: The Postlinear Approximation of General Relativity. \emph{Gen. Rel. Grav.} \textbf{13}, (1981). 
\bibitem{White} Whitehead, A.N. \emph{The Principle of Relativity}; Cambridge U. Press, 1922.
 \bibitem{Edding} A.S.  Eddington,  Nature  \textbf{113}, 192 (1924).
 \bibitem{Synge} J. L. Synge,
Orbits and Rays in the Gravitational Field of a Finite Sphere according to the Theory of A. N. Whitehead. Proc. of the Royal Soc. of London, \textbf{211}, 303  (1952).
 \bibitem{Will08}G.  Gibbons and  C.M. Will, Stud. Hist. Philos.~Mod.~Phys. \textbf{39}, 41 (2008).
 \bibitem{Fink}  D. Finkelstein,: Past-Future Asymmetry of the Gravitational Field of a Point Particle. \textit{Phys. Rev.} \textbf{1958} \textit{110} , p.965
 \bibitem{FStav} Y. Friedman and S. Stav,
 New metrics of a spherically symmetric gravitational field passing classical tests of General Relativity. 
  Europhys. Lett. \textbf{126}, 29001 (2019);  Erratum: Europhys. Lett. \textbf{127}, 19901 (2019) .
\bibitem{KerrSchild} Kerr, R.P. and Schild, A. Some algebraically degenerate solutions of Einstein's gravitational field equations. \textit{Proceedings of the Symposium on Applied Mathematics} \textbf{17}, 199-209, Providence: American Mathematical Society, 1965.

\bibitem{Kerr}  R. P. Kerr, Gravitational Field of a Spinning Mass as an Example of Algebraically Special Metrics. \textit{Phys. Rev. Lett..} \textbf{1963} \textit{11} (5): 237–238.
\bibitem{FUnify} Y. Friedman, 
A unifying physically meaningful relativistic action. 
Sci. Rep. \textbf{12}, 10843 (2022).    
\bibitem{ana} Y. Friedman and T. Scarr,  \emph{A Novel Approach to Relativistic Dynamics: Integrating Gravity, Electromagnetism and Optics,} Fundamental Theories of Physics 210 (Springer Nature, Switzerland, 2023)
  \bibitem{Jackson} J. D. Jackson  \emph{Classical Electrodynamics} (John Wiley \& Sons, 1998), 3rd ed.
 
\end{thebibliography}
\end{document}